\documentclass[3p,times]{elsarticle}

\usepackage{ecrc}

\volume{00}

\firstpage{1}

\journalname{Nuclear Physics A}

\runauth{Javier L. Albacete}

\jid{nupha}

\usepackage{amssymb}

\usepackage[figuresright]{rotating}

\begin{document}

\begin{frontmatter}



\dochead{}

\title{Testing the CGC in proton-lead collisions at the LHC}


\author{Javier L. Albacete}

\address{IPNO, Universit\'e Paris-Sud 11, CNRS/IN2P3, 91406 Orsay, France}

\begin{abstract}
I present a brief review of present CGC phenomenological applications and of the physics prospects for the forthcoming proton-lead run at the LHC
\end{abstract}

\begin{keyword}
Color Glass Condensate \sep High-density QCD   \sep LHC 

\end{keyword}

\end{frontmatter}

Besides its crucial role as a reference experiment to calibrate initial state effects in heavy ion collisions, the forthcoming p+Pb run at the LHC will provide access to kinematic regions never explored so far in
nuclear collisions and thus carries great potential for discovery of
new QCD phenomena on its own. In particular, the huge leap forward in
collision energy with respect to previous high energy electron-nucleus
or proton-nucleus experiments\footnote{The expected center of mass
  collision energy for the p+Pb run is 5~TeV, to be compared to a
  maximal energy of 200 GeV for RHIC d+Au collisions.} will probe the
nuclear wave function at values of Bjorken-$x$ smaller than ever
before. It is theoretically well established that at small enough
values of Bjorken-$x$ QCD enters a novel regime governed by large
gluon densities and non-linear coherence phenomena.  
The Color Glass Condensate (CGC) effective theory provides a
consistent framework to study QCD scattering at small-$x$ or high
collision energies (for a review see e.g.
\cite{Iancu:2003xm,Weigert:2005us}). It is based on three main
physical ingredients: First, high gluon densities correspond to strong
classical fields, which permit ab-initio first principles calculation
of ``wave functions'' at small $x$ through classical techniques. Next,
quantum corrections are incorporated via non-linear renormalization
group equations such as the B-JIMWLK hierarchy or, in the
large-$N_{c}$ limit, the BK
equation~\cite{Balitsky:1996ub,Kovchegov:1999yj} that describe the
evolution of the hadron wave function towards small $x$. The
non-linear, density-dependent terms in the CGC evolution equations are
ultimately related to unitarity of the theory and, in the appropriate
frame and gauge, can be interpreted as due to gluon recombination
processes that tame or {\it saturate} the growth of gluon densities
for modes with transverse momenta below a dynamically generated
scale known as the saturation scale, $Q_{s}(x)$. Finally, the
presence of strong color fields $\mathcal{A}\sim1/g$ leads to
breakdown of standard perturbative techniques to describe particle
production processes based on a series expansion in powers of the
strong coupling $g$. Terms of order $g\mathcal{A} \sim \mathcal{O}(1)$
need to be resummed to all orders. The CGC provides the tools to
perform such resummation although the precise prescription for the
resummation may vary from process to process or colliding system

While the CGC has been successfully applied to the description of different observables in different collision systems (from e+p to AA), the p+Pb run at the LHC will provide an excellent --and probably
in the near future unique-- possibility to disentangle the presently
inconclusive situation on the role of CGC effects and also to
distinguish among different approaches to describe high energy
scattering in nuclear reactions. On the one hand, the LHC shall bring us
closer to the limit of asymptotically high energy in which the CGC
formalism is developed, thus reducing theoretical uncertainties on its
applicability. Equivalently, the value of the saturation scale is
expected to be a factor $\sim2\div4$ times larger than at RHIC, so
saturation effects should be visible in a larger range of transverse
momenta, deeper into the perturbative domain. On the other hand, the
much extended reach in the LHC will allow measurements far from the
kinematic limit up to very forward rapidities, thus minimizing the
role of large-$x$ effects which obscured the interpretation of forward
RHIC data.

\section{Coherence effects in high-energy nuclear collisions}
\label{sec1}

A main lesson learnt from experimental data collected in (d)Au+Au and Pb+Pb collisions at RHIC and the LHC respectively is that bulk particle production in ion-ion collisions is very different from a simple superposition of nucleon-nucleon collisions. Such is evident in terms of the measured charged particle multiplicities, which exhibit a strong deviation from the scaling with the number of nucleon-nucleon collisions: $\frac{dN^{AA}}{d\eta}(\eta=0)\ll N_{coll} \frac{dN^{AA}}{d\eta}(\eta=0)$, and also from the non-trivial transverse momentum dependence of nuclear modification factors measured in d+Au collisions.  These observations lead to the conclusion that strong coherence effects among the constituent nucleons, or the relevant degrees of freedom at the sub-nucleon level, must be present during the collisions process. 

Indeed, and regardless of the question whether the CGC is the most suited
framework for their description there is broad consensus that
coherence effects are essential for the interpretation of present data
on heavy ion collisions. In fact most --if not all-- of the different
phenomenological approaches for the description of particle production
--both in the soft or hard sector--. On physical grounds, coherence phenomena are related to the presence of high gluon densities in the wave function of the colliding nuclei at small values of Bjorken-$x$\footnote{A rough estimate of the values of Bjorken-$x$ proven in a hadronic collision can be obtained using $2\rightarrow1$ kinematics $x_{1(2)}\sim \frac{p_t}{\sqrt{s}}\,e^{\pm y}$, with $\sqrt{s}$, $p_{t}$ and $y$ being the collision energy and the transverse momentum and rapidity of the produced hadron respectively.} 
While a detailed discussion of the different prescriptions found in the literature to account for coherence effects is beyond the scope of this brief review, one can identify in different models coherence effects at the level of the wave function and also at the level of primary particle production, sketched in Fig. 1 (left)~\footnote{Here we just mention here a few well known
  examples; a rather exhaustive compilation of phenomenological works
  to for the description of particle production in HIC can be found in
  e.g \cite{Abreu:2007kv}.}. To the first category correspond the nuclear shadowing (in a partonic language) or the percolation and string fusion (in non-perturbative approaches). In both cases, when different constituents, whichever the degrees of freedom chosen are, overlap in phase space according to some geometric criterium, recombination of such constituents happen, thus reducing the total number of scattering centers --gluons-- entering the collision process. Similar phase-space arguments motivate the implementation of energy-dependent cut-offs to regulate independent particle production from different sources, normally a working hypothesis in most Monte Carlo event generators for heavy ion collisions. In Fig 1 (right) we show the collision energy dependence of the transverse momentum cutoff that separates {\it hard} from {\it soft} particle production in several event generators for p+p (PYTHIA) and A+A (HYDJETY and HIJING) collisions. Its strong rise with increasing energy signals the increasing importance of collective effects in particle production processes. Also frequent in the literature are the resummation of multiple scatterings, either in a coherent higher twist formalism or in a incoherent {\it Glauber like} formalism. Finally, the modification of parton distribution functions to allow for the presence of a --normally energy dependent-- intrinsic transverse momentum constitute other common practice in phenomenological works.  
All these ingredients are akin, at least at a conceptual level, to those dynamically built in the CGC, although they are formulated in very different ways. Thus the debate is now which theoretical framework is most suited for their description: whether the CGC (at its present degree of accuracy!) or alternative approaches, typically rooted in the standard collinear factorization framework. 

\begin{figure}[htb]
\begin{center}
\includegraphics[width=0.47\textwidth]{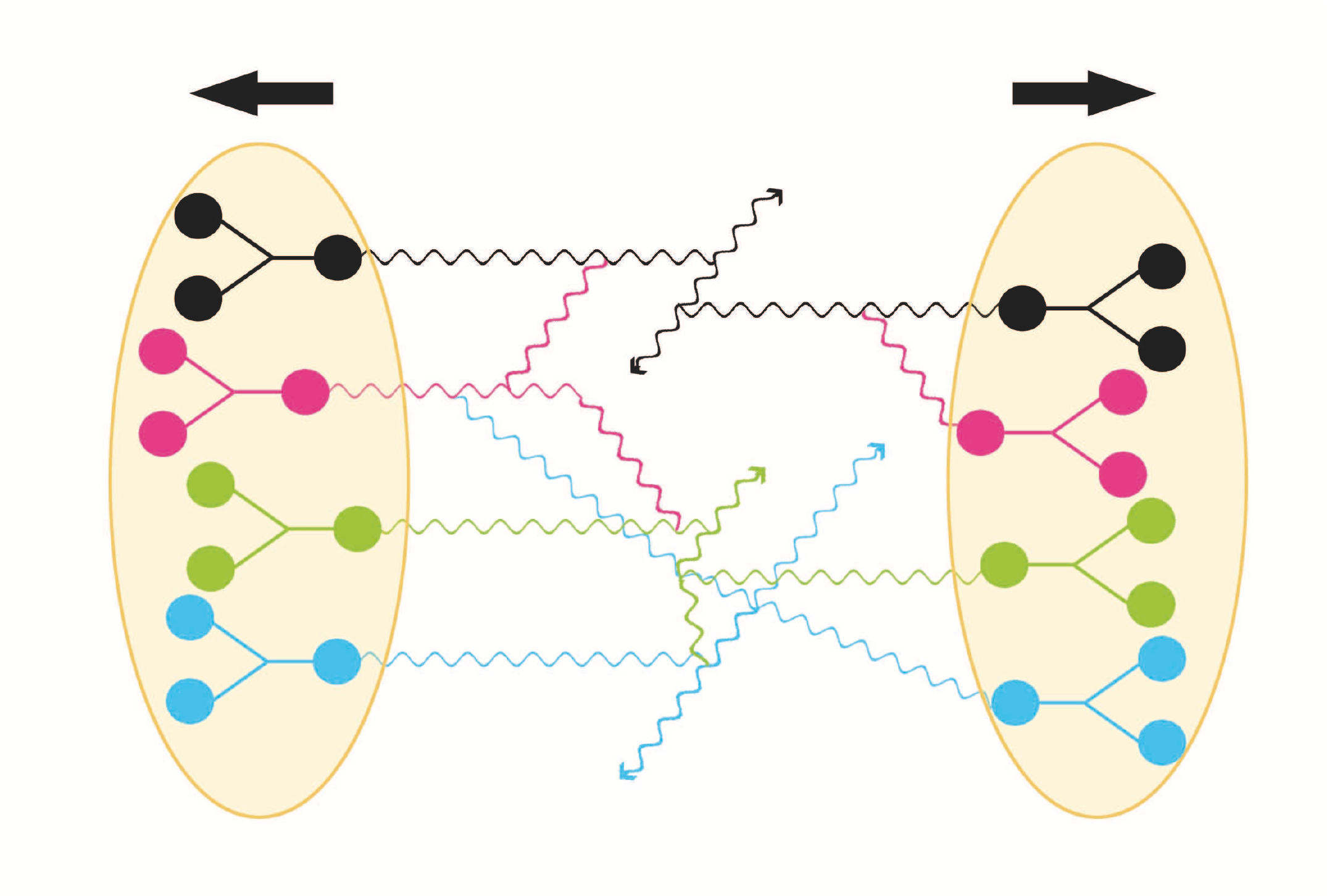}
\includegraphics[width=0.51\textwidth]{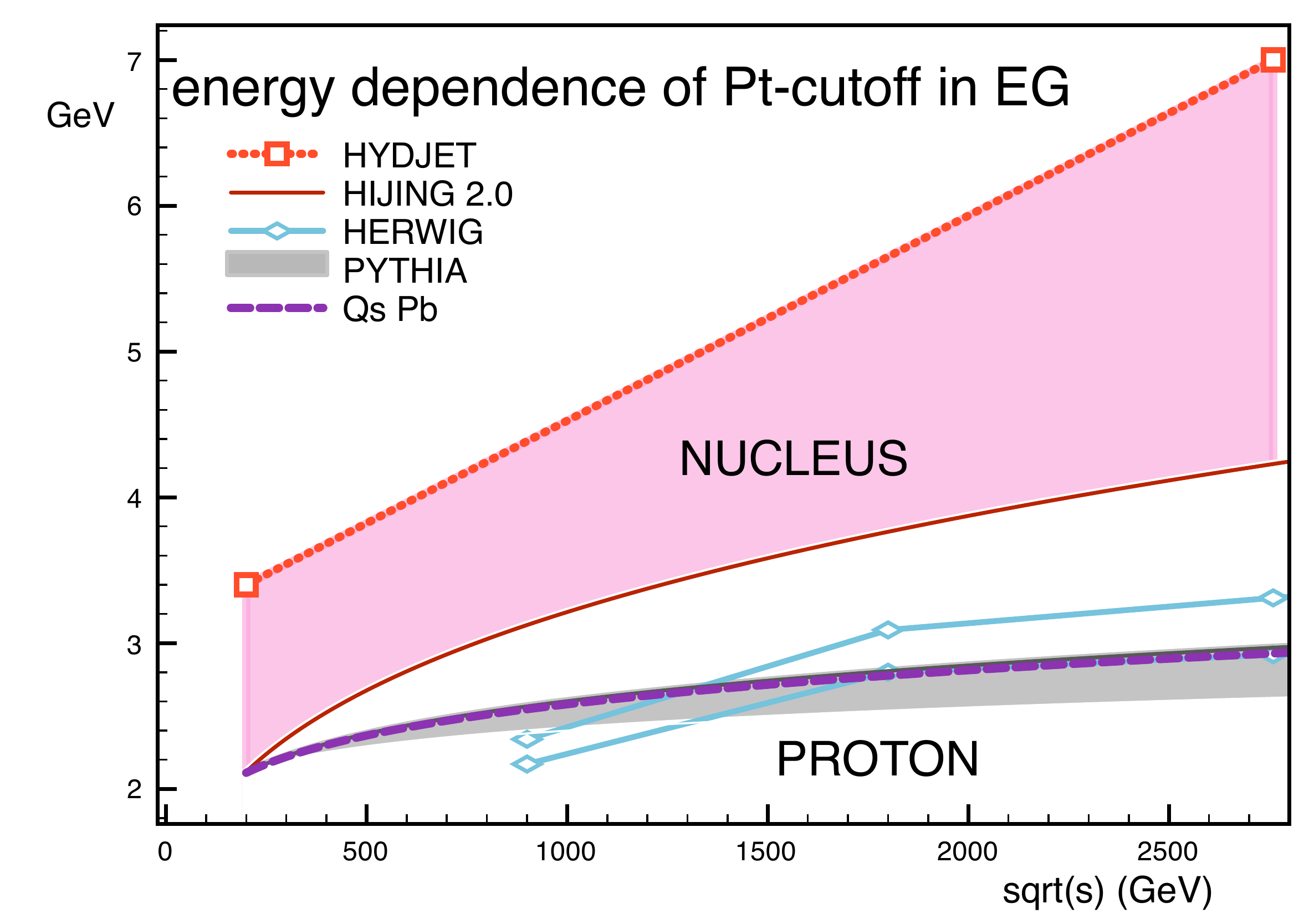}
\end{center}
\vspace*{-0.5cm}
\caption[a]{Left: Schematic representation of coherence effects in heavy ion collisions. Right: Energy-dependence of the transverse momentum cutoff in MC event generators for proton-proton collisions (two different PYTHIA tunes) and heavy ion collisions (HYDJET and HIJING). Also shown the characteristic value of the saturation scale $Q_{s}\sim \sqrt{s}^{0.15}\,$. }
\label{coher}
\end{figure}

\section{State-of-the-art of the CGC formalism and its phenomenological applications}
\label{sec2}

Briefly, it can be said that the CGC is now entering the next-to-leading (NLO) order era. Actually, the kernel of the BK and JIMWLK evolution equations are now known to NLO accuracy \cite{Balitsky:2008zz} or, also, to running coupling accuracy through the resummation of a partial subset of NLO diagrams \cite{Kovchegov:2006vj,Balitsky:2006wa,Albacete:2007yr}. In a similar fashion, the following calculations concerning particle production processes have recently become available at NLO accuracy: photon impact factors in deep inelastic scattering~\cite{Balitsky:2012bs}, full NLO~\cite{Chirilli:2011km} and {\it inelastic}~\cite{Altinoluk:2011qy} contributions to hybrid calculation of single inclusive particle production in dilute-dense scattering, running coupling corrections to the $k_{t}$-factorization formula~\cite{Horowitz:2010yg} and a proof of factorization of multiparticle production processes at NLO~\cite{Gelis:2008rw}. Furthermore, our knowledge of exclusive particle production and multi-particle correlations (di-hadrons, hadron-photon etc) has been advanced significantly through a series of recent works that establish the precise relation between the $n$-point functions of the nuclear wave function and the observables of interest~\cite{Dominguez:2011wm, Dumitru:2011zz}. Thus, and despite the intrinsic technical difficulty of higher order calculations in the CGC --one should recall that they are performed by expanding on a strong background color field-- progress on the theoretical side has been steadily delivered over the last years.

Notwithstanding the progress brought by these works, the CGC framework is still far from the degree of sophistication and accuracy that characterizes standard pQCD methods based on the collinear factorization formalism and DGLAP evolution equations. The reasons for this are manifold: First, the new theory tools discussed above have not been yet fully implemented in a systematic way in phenomenological applications. Current descriptions of data combine NLO with LO theory ingredients in a somehow uncontrolled theoretical scheme. For instance, the --arguably-- most ambitious effort so far to determine the universal properties of nuclear and proton wave functions at small-$x$ in a systematic way is provided by the global AAMQS fits to 
e+p data\cite{Albacete:2009fh,Albacete:2010sy} and their extension to the nuclear case through Monte Carlo methods~\cite{Albacete:2010ad}. The AAMQS fits are rely on the use of the running coupling BK equation to describe the small-$x$ evolution of the 2-point function,  but then use the LO dipole formalism to calculate the $\gamma^{*}$-proton cross section. Related issues obscure the successful phenomenological  descriptions of data in single~\cite{Albacete:2010bs} and double~\cite{Albacete:2010pg} inclusive production in p+p and d+Au collisions or the widespread use of $k_{t}$-factorization in models for total multiplicities, known not to be valid in the case of {\it dense-dense} scattering.
Most of the NLO tools being available, the main obstacle to fix such insufficiencies of present phenomenological works is the difficulty of their technical implementation. NLO and exclusive particle production calculations typically require the knowledge of the $x$-dependence of $n$-point functions beyond the 2-point function. Hence, the use of the (running coupling) BK-equation for the 2-point function, easy to solve numerically, is no longer sufficient and solutions of the full B-JIMWLK hierarchy of coupled equations --more demanding numerically-- are needed. A promising analytic method to solve B-JIMWLK equations through a Gaussian approximation that allows to obtain arbitrary $n$-point functions in terms of the 2-point one has been proposed recently~\cite{Iancu:2011ns}, and its practical implementation would substantially reduce the uncertainties progress related to calculations on di-hadron correlations.

Other factor that blurs the predictive power of the CGC framework is the paucity of experimental data on high-energy (equivalently, small-$x$) nuclear reactions. Such information is needed to constrain the non-perturbative parameters of the theory, like the initial conditions for the evolution at some initial scale ($x_{0}$ tipically taken to be $=10^{-2}$ in practical applications) or the impact parameter dependence of the nuclear unintegrated nuclear distributions. 
Lacking such information, some degree of modeling is unavoidable. Mean field approaches for the description of the nuclear geometry have been recently superseded by Monte Carlo methods as the MC-KLN or MC-rcBK ones ~\cite{Drescher:2007ax, Albacete:2010ad}, where the position of nucleons in the transverse plane are treated as a random variable, thus allowing to account for geometry fluctuations in the collision process.

Keeping in mind the present limitations of CGC phenomenological works discussed in this section, let me now briefly review current CGC predictions for different observables at the LHC in the next sections.

\section{Multiplicities}
\label{sec3}
The CGC offers a very economical description of the integrated hadron multiplicities produced in heavy ion collisions. Based solely in dimensional analysis, the number of particles produced per unit of transverse area rises proportional to the saturation scale (in symmetric collisions)
\begin{equation}
\left.\frac{dN}{d\eta d^{2}b}\right |_{\eta=0}\sim Q_{s}^{2}(\sqrt{s},b).  
\end{equation}
Such approximate expression can be realized either via solutions of the classical equation of motion in the presence of sources (projectile and target) or --less rigorously for A+A collisions-- through the use of $k_{t}$-factorization. It accounts well for two of the most remarkable features observed in RHIC d+Au and Au+Au data as well as in LHC Pb+Pb data: {\it i)} Approximate factorization of the collision energy and collision centrality of total multiplicities and  {\it ii)} Power-law dependence on integrated multiplicities on collision energy. In Fig 2 we show the predictions for the rapidity dependence of hadron multiplicities from different CGC models rooted in $k_{t}$-factorization, where particle production is given by the convolution of the unintegrated gluon distributions of projectile and target. These models differ mainly  in their input for the nuclear unintegrated gluon distributions and in the treatment of the geometry dependence --either Monte Carlo methods or mean field approaches-- but also in the implementation details like the IR regularization or the rapidity to pseudo-rapidity Jacobian or the modeling of the large-$x$ component of the corresponding wave functions (hence the large deviations among them at more forward/backward rapidities). They have been tested against d+Au RHIC data and then extrapolated to LHC energies without further adjustments. Overall they predict a charged hadron multiplicity 
\begin{equation}
\left.\frac{dN^{p+Pb}_{ch}(\sqrt{s}=5 \,TeV)}{d\eta} \right |_{\eta=0}\approx 17\pm2\,.  
\end{equation}

\begin{figure}[htb]
\begin{center}
\includegraphics[width=0.65\textwidth]{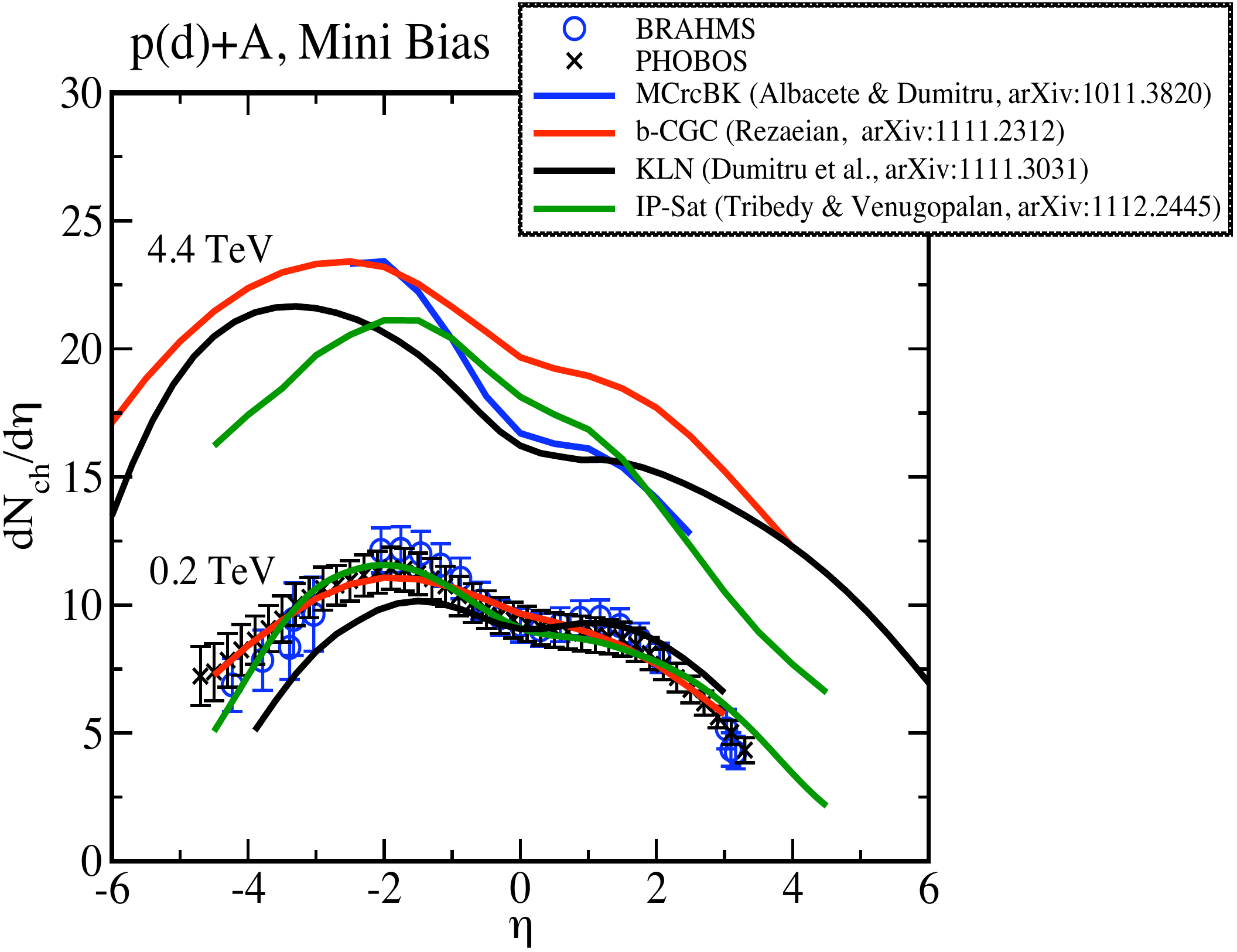}
\end{center}
\vspace*{-0.5cm}
\caption[a]{ A compilation of CGC-based predictions for the repidity distribution of integrated hadron multiplicities in p+Pb collisions at the LHC: MCrcBK~\cite{Albacete:2010ad}, b-CGC~\cite{Rezaeian:2011ia}, MC-KLN~\cite{Dumitru:2011wq} and IP-Sat~\cite{Tribedy:2011aa}. Fig courtesy of A. Rezaeian.   }
\label{mult}
\end{figure}

\section{Nuclear modification factors}
\label{sec4}

An observable that has centered much of the discussion on the relevance of CGC physics in heavy ion collisions is the nuclear modification factor measured in d+Au collisions at RHIC. At mid-rapidity RHIC data show a moderate enhancement of single inclusive particle production tipically atributed to semi-classical multiple scatterings, whereas they are continuously depleted at more forward rapidities. A good quantitative description of this forward suppression is possible within the CGC~\cite{Albacete:2010bs}, where it relates to the non-linear evolution of the nuclear ugd. However, RHIC forward data falls very close to the kinematic limit, where large-$x$ effects such as energy loss --neglected in the in the CGC approach-- may also be relevant. Indeed, there are alternative explanations of data where energy loss is the main dynamical ingredient (see e.g~\cite{Kopeliovich:2005ym}). Moreover, the CGC description of the most forward data on neutral pions performed by the STAR collaboration requires the use of $K$-factors smaller than unity, hinting  at the relevance of large-$x$ effects. 

Although more exclusive observables like di-hadron or hadron-photon
correlations are expected to better discriminate between different
approaches, a first test for models of particle production in HIC
shall come from data on inclusive multiplicities and single particle
distributions as they are much easier to obtain experimentally. 
Generic arguments based on kinematics suggest that one should expect a similar suppression of the nuclear modification factors in LHC p+Pb collisions at mid-rapidity as the one observed in forward RHIC d+Au data, since in both cases the nuclear wave function is probed at similar values of Bjorken-$x$. While this argument can be a somewhat misleading (the proximity of RHIC forward data to the kinematic limit limits strongly the transverse momentum of the gluons probed in the nucler ugd), it is approximately realized in the different CGC-predictions. Fig 3 show the predictions for $R_{pPb}$ at rapidity 0 (left)  and $2$ (right) based from the rcBK MC~\cite{prep}  and IP-Sat~\cite{Tribedy:2011aa} models. They predict a moderate suppression (similar to the one observed at RHIC $\sim0.6\pm0.7$ at moderate to small transverse momentum $p_{t}\lesssim 3\pm4$ GeV, and a smooth approach to unity at larger transverse momentum. The error bands in the MC-rcBK calculation originate from the use of different initial conditions to solve the running coupling BK evolution and also from the variation of factorization scales and fragmentation functions in the $k_{t}$-factorization formula they rely upon. Also, it should be note that the MC-rcBK predictions presented here correspond to the {\it quenched} approximation for the nuclear geometry. The details of this calculation will be presented in ~\cite{prep}. Remarkably, the CGC predictions for mid-rapidity overlap strongly with those obtained within the collinear factorization framework using the EPS09 parametrization for the nPDF's (results from~\cite{QuirogaArias:2010wh}). One then concludes that only this observable does not suffice to discriminate among different approaches to particle production in high energy heavy ion collisions. 
The differences between CGC and collinear factorization formalisms start becoming apparent as one moves to more forward rapidities.  CGC models predict a faster onset of the suppression at higher values of transverse momentum with respect to he EPS09 results. The weak rapidity dependence of EPS09 results can be traced back to the flatness of the gluon modification factor at small-$x$ in this parametrization. 
Thus, a rapidity scan of the nuclear modification factors would offer a much larger discriminating power.
  
CGC predictions at intermediate values of rapidity suffer of a systematic uncertainty:
In the CGC there are two distinct but related approaches to hadron production in
high energy asymmetric collisions.  Particle production processes in the central rapidity region probe the
wave functions of both projectile and target at small values of
$x$. Here, one may employ the $k_t$-factorization formalism where both
the projectile and target are characterized in terms of their rcBK
evolved unintegrated gluon distributions (UGDs). 
However, at more forward rapidities, the proton is probed at
larger values of $x$ while the target nucleus is shifted deeper into
the small-$x$ regime. Here, $k_{t}$-factorization fails to
grasp the dominant contribution to the scattering process. Rather, the
{\it hybrid} formalism proposed in ref.~\cite{Dumitru:2005gt}. In the hybrid formalism the
large-$x$ degrees of freedom of the proton are described in terms of
usual parton distribution functions (PDFs) of collinear factorization
which satisfy the momentum sum rule exactly and which exhibit a scale
dependence given by the DGLAP evolution equations. On the other hand,
the small-$x$ glue of the nucleus is still described in terms of its
UGD. The corresponding limits of applicability of each formalism --equivalently the precise value
of $x$ at which one should switch from one to the other-- have only
been estimated on an empirical basis, and in practice it is taken to be $x_{0}\sim 10^{-2}$. 

The Leading Order hybrid formalism yield a stronger suppression that the $k_{t}$-factorization one (see Fig 2 right).  Recently the hybrid formalism has been improved through the
calculation of inelastic contributions and full NLO corrections that may become important at
high transverse momentum~\cite{Altinoluk:2011qy}. The inelastic terms were recently implemented in phenomenological work~\cite{JalilianMarian:2011dt}. There it was observed that the effect of the inelastic corrections is to slightly increase the value of the nuclear modification factors with respect to the LO result, thus bringing them closer to the results obtained within the $k_{t}$-factorization framework. A full phenomenological implementation of the hybrid and $k_{t}$-factorization at NLO is necessary to better asses this systematic uncertainty.  

At a strictly qualitative level, two generic features of CGC predictions can be highlighted:
\begin{itemize}
\item[a)]: Dissapearence of the Cronin peak in $R_{pPb}$ at central rapidities. 
\item[b)]: Stronger suppression at forward rapidities:  $R^{pPb}(y_1,p_{t})>R^{pPb}(y_2,p_{t})$ for $0>y_{1}<y_{2}$.
\end{itemize}
These two features of CGC predictions originate from generic properties of the non-linear small-$x$ evolution, regardless of the degree of accuracy of the evolution kernel, NLO or LO etc. Therefore, the persistence of the Cronin peak in $R_{pPb}$ at the LHC would be very difficult to accommodate in the CGC framework. Similarly the continuos depletion of nuclear modification factors reflects the relative enhancement of non-linear correction to the evolution of nuclear wave function with respect to that of a proton.

\begin{figure}[htb]
\begin{center}
\includegraphics[width=0.49\textwidth]{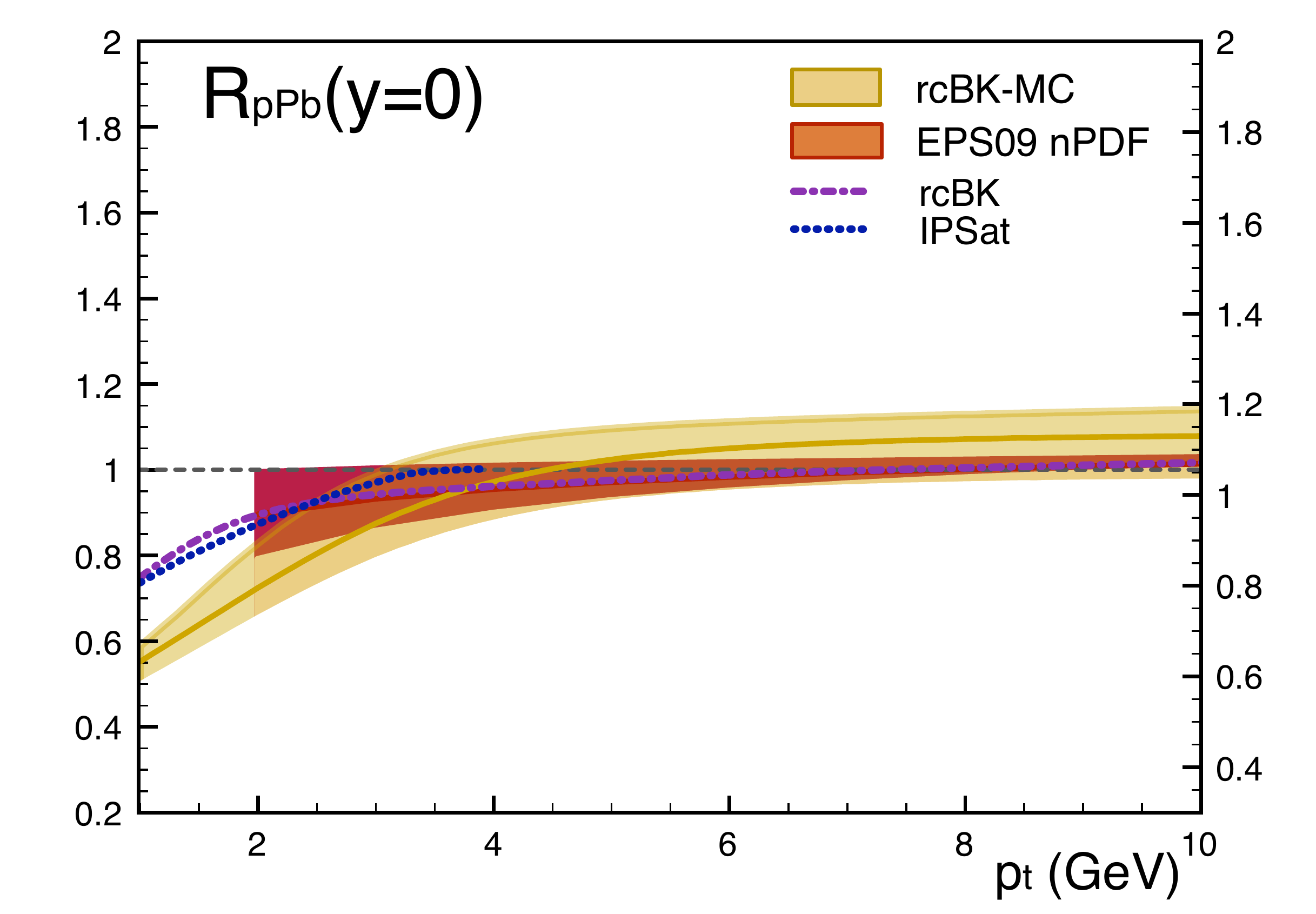}
\includegraphics[width=0.49\textwidth]{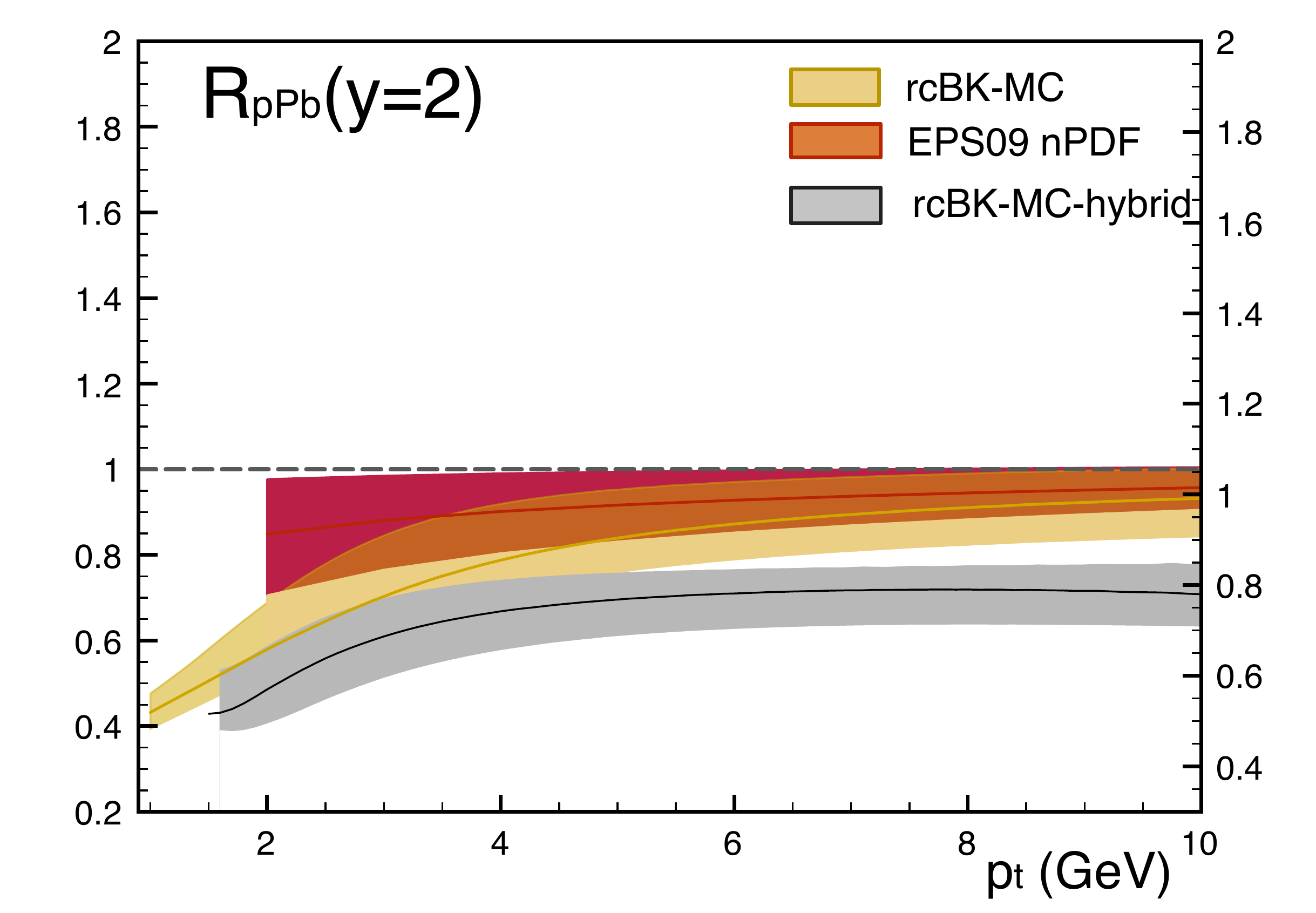}
\end{center}
\vspace*{-0.5cm}
\caption[a]{MC-rcBK~\cite{prep} and IP-Sat~\cite{Tribedy:2011aa}(dashed lines in left plot) predictions for the nuclear modification factors for neutral pions at the LHC at rapidities 0 (left) and 2 (right). The dashed lines The yellow or grey bands shown at $y=2$ correspond to results obtained within the $k_{t}$-factorization or hybrid formalism respectively. Also shown in red the predictions corresponding to the EPS09 nPDF set within collinear factorization~\cite{QuirogaArias:2010wh} (courtesy of P. Quiroga-Arias). }
\label{rpa}
\end{figure}

\section{Two-particle correlations}
\label{sec5}
The study of forward di-hadron correlations in d+Au azimuthal correlations at RHIC has provided the most solid indication for the relevance of CGC effects in data so far. The disappearance of the away side peak --also dubbed {\it monojet} production-- in more central collisions (and its persistence in more peripheral collisions and p+p collisions) can be related to 
the interplay between the transverse momenta of the produced hadrons and the one acquired during the
interaction with the nucleus. In the CGC approach the interaction with the nucleus is realized
in a fully coherent way, and the momentum broadening
is parametrically controlled by the $x$-dependent saturation
scale of the nucleus. The latter, in turn, is described
by means of the rcBK equation. A first semi-quantitative description of the data was provided in~\cite{Albacete:2010pg}. This work neglected some leading in $N_{c}$ terms which calculation demands knowledge of higher $n$-point functions and also the contribution of the gluon channel to the production process. These caveats were partially fixed in a later work~\cite{Stasto:2011ru}, where the small momentum imbalance approximation was used. Moreover, the role of multiparton interactions that may enhance the contribution of the uncorrelated component of the double inclusive cross section (specially in the forward region)~\cite{Strikman:2010bg} has not been fully explored in present CGC calculations. Nevertheless, the description of data is good in these two works is good, see Fig 4. Recently, another description of data based in a higher-twist calculation that also includes nuclear shadowing and cold nuclear matter energy loss has become available~\cite{Kang:2011bp}. A full CGC analyses of data on di-hadron correlations including the missing ingredients in previous CGC analyses is now underway~\cite{Lappi:2012xe}. It should serve as the reference to generate precise quantitative predictions for p+Pb collisons at the LHC. So far, qualitative expectations indicate that analogous suppression of azimuthal correlations should be observed in at the LHC. Generically the strength of the decorrelation is expected to be stronger with: {\it i)} increasing rapidity of the produced pair; {\it ii)} increasing collision centrality and {\it iii)} decreasing transverse momentum of the trigger and associated particle. 

It has been recently proposed that hadron-photon~\cite{JalilianMarian:2012bd} and hadron-dileptons~\cite{Stasto:2012ru} correlations may exhibit similar azimuthal structure. These observables offer the advantage that they can be computed in terms of only the rcBK-evolved 2-point function. Although their experimental determination may be more complicated, their measurement at the LHC would provide additional constraints to determine the underlying dynamics of multiparticle production in high-density QCD scattering.

\begin{figure}[htb]
\begin{center}
\includegraphics[width=0.5\textwidth]{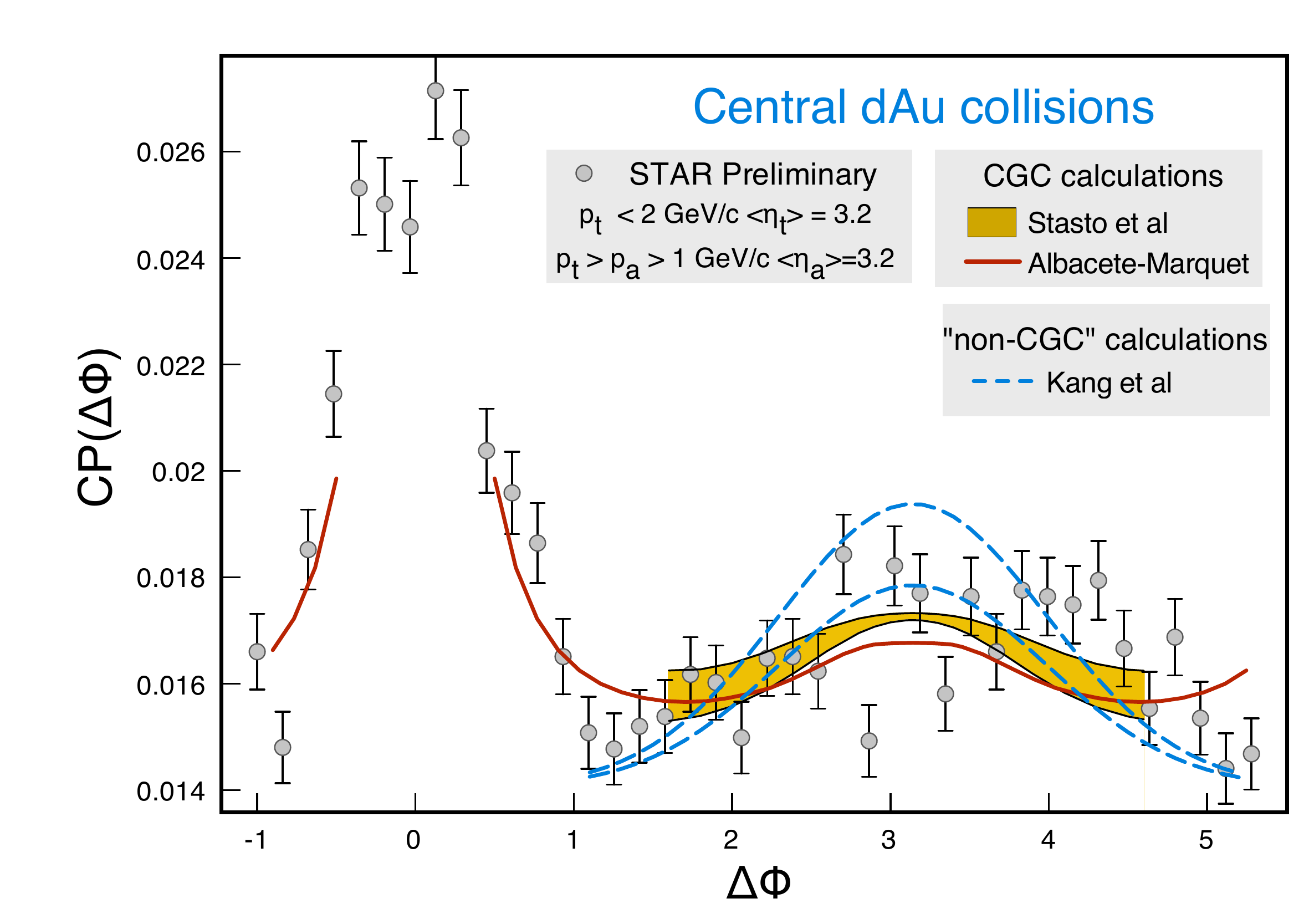}
\end{center}
\vspace*{-0.5cm}
\caption[a]{Comparison of STAR preliminary data~\cite{Braidot:2010zh} for the Coincident Probability between pairs of hadrons as a function of the relative azimuthal angle in d+Au collisions at RHIC. The theoretical results correspond to two CGC-based calculations\cite{Albacete:2010pg} and \cite{Stasto:2011ru} and a {\it higher-twist} one~\cite{Kang:2011bp}.  }
\label{corr}
\end{figure}

\section{Conclusions}
\label{sec5}

In summary, while it is fair to say that a large number of observables in different systems --from e+p to A+A collisions-- that probe the small-$x$ component of the wave function of the projectile or target find their natural interpretation in terms of high gluonic densities and also a good quantitative description in terms of CGC-based calculations, no conclusive claim for the observation saturation physics can be performed yet.
Important steps have been taken over the last years in promoting the CGC framework to a predictive and quantitative phenomenological tool. Such has been possible through the systematic implementation of global fit and Monte Carlo methods and, more importantly, through an intense theoretical work in the determination of higher order corrections to the formalism, including running coupling corrections to non-linear evolution equations and also to particle production processes. Nevertheless, this program is far for complete and there is still a large margin for improvement in the CGC phenomenological works. 
The p+Pb data will provide precious information to sharpen the CGC quantitative tool. First of all it will provide empiric information needed to constrain the non-perturbative parameters of the theory. Next it will allow to test the generic CGC predictions and also whether the present degree of accuracy of the CGC effective theory is sufficient to quantitatively describe data.

\section*{Acknowlegments}
I would like to thank the organizers for their invitation to this very interesting conference. My research is supported by a fellowship from the Th\'eorie LHC France initiative funded by the IN2P3.





\bibliographystyle{elsarticle-num}

\begin{thebibliography}{10}
\expandafter\ifx\csname url\endcsname\relax
  \def\url#1{\texttt{#1}}\fi
\expandafter\ifx\csname urlprefix\endcsname\relax\def\urlprefix{URL }\fi
\expandafter\ifx\csname href\endcsname\relax
  \def\href#1#2{#2} \def\path#1{#1}\fi

\bibitem{Iancu:2003xm}
E.~Iancu, R.~Venugopalan, The color glass condensate and high energy scattering
  in {QCD}\href {http://arxiv.org/abs/hep-ph/0303204}
  {\path{arXiv:hep-ph/0303204}}.

\bibitem{Weigert:2005us}
H.~Weigert, Evolution at small x: The color glass condensate, Prog. Part. Nucl.
  Phys. 55 (2005) 461--565.
\newblock \href {http://arxiv.org/abs/hep-ph/0501087}
  {\path{arXiv:hep-ph/0501087}}.

\bibitem{Balitsky:1996ub}
I.~Balitsky, Operator expansion for high-energy scattering, Nucl. Phys. B463
  (1996) 99--160.
\newblock \href {http://arxiv.org/abs/hep-ph/9509348}
  {\path{arXiv:hep-ph/9509348}}.

\bibitem{Kovchegov:1999yj}
Y.~V. Kovchegov, Small-x {$F_2$} structure function of a nucleus including
  multiple pomeron exchanges, Phys. Rev. D60 (1999) 034008.
\newblock \href {http://arxiv.org/abs/hep-ph/9901281}
  {\path{arXiv:hep-ph/9901281}}.

\bibitem{Abreu:2007kv}
N.~Armesto, (ed.~), et~al., {Heavy Ion Collisions at the LHC - Last Call for
  Predictions}, J. Phys. G35 (2008) 054001.
\newblock \href {http://arxiv.org/abs/0711.0974} {\path{arXiv:0711.0974}},
  \href {http://dx.doi.org/10.1088/0954-3899/35/5/054001}
  {\path{doi:10.1088/0954-3899/35/5/054001}}.

\bibitem{Balitsky:2008zz}
I.~Balitsky, G.~A. Chirilli, {Next-to-leading order evolution of color
  dipoles}, Phys. Rev. D77 (2008) 014019.
\newblock \href {http://arxiv.org/abs/0710.4330} {\path{arXiv:0710.4330}},
  \href {http://dx.doi.org/10.1103/PhysRevD.77.014019}
  {\path{doi:10.1103/PhysRevD.77.014019}}.

\bibitem{Kovchegov:2006vj}
Y.~Kovchegov, H.~Weigert, {Triumvirate of Running Couplings in Small-$x$
  Evolution}, Nucl. Phys. {\bf A} 784 (2007) 188--226.
\newblock \href {http://arxiv.org/abs/hep-ph/0609090}
  {\path{arXiv:hep-ph/0609090}}.

\bibitem{Balitsky:2006wa}
I.~I. Balitsky, {Quark Contribution to the Small-$x$ Evolution of Color
  Dipole}, Phys. Rev. D 75 (2007) 014001.
\newblock \href {http://arxiv.org/abs/hep-ph/0609105}
  {\path{arXiv:hep-ph/0609105}}.

\bibitem{Albacete:2007yr}
J.~L. Albacete, Y.~V. Kovchegov, Solving high energy evolution equation
  including running coupling corrections, Phys. Rev. D75 (2007) 125021.
\newblock \href {http://arxiv.org/abs/arXiv:0704.0612 [hep-ph]}
  {\path{arXiv:arXiv:0704.0612 [hep-ph]}}.

\bibitem{Balitsky:2012bs}
I.~Balitsky, G.~A. Chirilli, {Photon impact factor and $k_T$-factorization for
  DIS in the next-to-leading order}\href {http://arxiv.org/abs/1207.3844}
  {\path{arXiv:1207.3844}}.

\bibitem{Chirilli:2011km}
G.~A. Chirilli, B.-W. Xiao, F.~Yuan, {One-loop Factorization for Inclusive
  Hadron Production in $pA$ Collisions in the Saturation Formalism},
  Phys.Rev.Lett. 108 (2012) 122301.
\newblock \href {http://arxiv.org/abs/1112.1061} {\path{arXiv:1112.1061}},
  \href {http://dx.doi.org/10.1103/PhysRevLett.108.122301}
  {\path{doi:10.1103/PhysRevLett.108.122301}}.

\bibitem{Altinoluk:2011qy}
T.~Altinoluk, A.~Kovner, {Particle Production at High Energy and Large
  Transverse Momentum - 'The Hybrid Formalism' Revisited}, Phys.Rev. D83 (2011)
  105004.
\newblock \href {http://arxiv.org/abs/1102.5327} {\path{arXiv:1102.5327}},
  \href {http://dx.doi.org/10.1103/PhysRevD.83.105004}
  {\path{doi:10.1103/PhysRevD.83.105004}}.

\bibitem{Horowitz:2010yg}
W.~Horowitz, Y.~V. Kovchegov, {Running Coupling Corrections to High Energy
  Inclusive Gluon Production}, Nucl.Phys. A849 (2011) 72--97.
\newblock \href {http://arxiv.org/abs/1009.0545} {\path{arXiv:1009.0545}},
  \href {http://dx.doi.org/10.1016/j.nuclphysa.2010.10.014}
  {\path{doi:10.1016/j.nuclphysa.2010.10.014}}.

\bibitem{Gelis:2008rw}
F.~Gelis, T.~Lappi, R.~Venugopalan, {High energy factorization in
  nucleus-nucleus collisions}, Phys. Rev. D78 (2008) 054019.
\newblock \href {http://arxiv.org/abs/0804.2630} {\path{arXiv:0804.2630}},
  \href {http://dx.doi.org/10.1103/PhysRevD.78.054019}
  {\path{doi:10.1103/PhysRevD.78.054019}}.

\bibitem{Dominguez:2011wm}
F.~Dominguez, C.~Marquet, B.-W. Xiao, F.~Yuan, {Universality of Unintegrated
  Gluon Distributions at small x}, Phys.Rev. D83 (2011) 105005.
\newblock \href {http://arxiv.org/abs/1101.0715} {\path{arXiv:1101.0715}},
  \href {http://dx.doi.org/10.1103/PhysRevD.83.105005}
  {\path{doi:10.1103/PhysRevD.83.105005}}.

\bibitem{Dumitru:2011zz}
A.~Dumitru, J.~Jalilian-Marian, E.~Petreska, {Two-gluon correlations and
  initial conditions for small-x evolution}, Phys.Rev. D84 (2011) 014018.
\newblock \href {http://arxiv.org/abs/1105.4155} {\path{arXiv:1105.4155}},
  \href {http://dx.doi.org/10.1103/PhysRevD.84.014018}
  {\path{doi:10.1103/PhysRevD.84.014018}}.

\bibitem{Albacete:2009fh}
J.~L. Albacete, N.~Armesto, J.~G. Milhano, C.~A. Salgado, {Non-linear QCD meets
  data: A global analysis of lepton- proton scattering with running coupling BK
  evolution}, Phys. Rev. D80 (2009) 034031.
\newblock \href {http://arxiv.org/abs/0902.1112} {\path{arXiv:0902.1112}},
  \href {http://dx.doi.org/10.1103/PhysRevD.80.034031}
  {\path{doi:10.1103/PhysRevD.80.034031}}.

\bibitem{Albacete:2010sy}
J.~L. Albacete, N.~Armesto, J.~G. Milhano, P.~Quiroga~Arias, C.~A. Salgado,
  {AAMQS: A non-linear QCD analysis of new HERA data at small-x including heavy
  quarks}, Eur.Phys.J. C71 (2011) 1705, * Temporary entry *.
\newblock \href {http://arxiv.org/abs/1012.4408} {\path{arXiv:1012.4408}},
  \href {http://dx.doi.org/10.1140/epjc/s10052-011-1705-3}
  {\path{doi:10.1140/epjc/s10052-011-1705-3}}.

\bibitem{Albacete:2010ad}
J.~L. Albacete, A.~Dumitru, {A model for gluon production in heavy-ion
  collisions at the LHC with rcBK unintegrated gluon densities}\href
  {http://arxiv.org/abs/1011.5161} {\path{arXiv:1011.5161}}.

\bibitem{Albacete:2010bs}
J.~L. Albacete, C.~Marquet, {Single Inclusive Hadron Production at RHIC and the
  LHC from the Color Glass Condensate}, Phys. Lett. B687 (2010) 174--179.
\newblock \href {http://arxiv.org/abs/1001.1378} {\path{arXiv:1001.1378}},
  \href {http://dx.doi.org/10.1016/j.physletb.2010.02.073}
  {\path{doi:10.1016/j.physletb.2010.02.073}}.

\bibitem{Albacete:2010pg}
J.~L. Albacete, C.~Marquet, {Azimuthal correlations of forward di-hadrons in
  d+Au collisions at RHIC in the Color Glass Condensate}, Phys.Rev.Lett. 105
  (2010) 162301.
\newblock \href {http://arxiv.org/abs/1005.4065} {\path{arXiv:1005.4065}},
  \href {http://dx.doi.org/10.1103/PhysRevLett.105.162301}
  {\path{doi:10.1103/PhysRevLett.105.162301}}.

\bibitem{Iancu:2011ns}
E.~Iancu, D.~Triantafyllopoulos, {Higher-point correlations from the JIMWLK
  evolution}, JHEP 1111 (2011) 105.
\newblock \href {http://arxiv.org/abs/1109.0302} {\path{arXiv:1109.0302}},
  \href {http://dx.doi.org/10.1007/JHEP11(2011)105}
  {\path{doi:10.1007/JHEP11(2011)105}}.

\bibitem{Drescher:2007ax}
H.-J. Drescher, Y.~Nara, {Eccentricity fluctuations from the Color Glass
  Condensate at RHIC and LHC}, Phys. Rev. C76 (2007) 041903.
\newblock \href {http://arxiv.org/abs/0707.0249} {\path{arXiv:0707.0249}},
  \href {http://dx.doi.org/10.1103/PhysRevC.76.041903}
  {\path{doi:10.1103/PhysRevC.76.041903}}.

\bibitem{Rezaeian:2011ia}
A.~H. Rezaeian, {Charged particle multiplicities in pA interactions at the LHC
  from the Color Glass Condensate}, Phys.Rev. D85 (2012) 014028.
\newblock \href {http://arxiv.org/abs/1111.2312} {\path{arXiv:1111.2312}},
  \href {http://dx.doi.org/10.1103/PhysRevD.85.014028}
  {\path{doi:10.1103/PhysRevD.85.014028}}.

\bibitem{Dumitru:2011wq}
A.~Dumitru, D.~E. Kharzeev, E.~M. Levin, Y.~Nara, {Gluon Saturation in $pA$
  Collisions at the LHC: KLN Model Predictions For Hadron Multiplicities},
  Phys.Rev. C85 (2012) 044920.
\newblock \href {http://arxiv.org/abs/1111.3031} {\path{arXiv:1111.3031}},
  \href {http://dx.doi.org/10.1103/PhysRevC.85.044920}
  {\path{doi:10.1103/PhysRevC.85.044920}}.

\bibitem{Tribedy:2011aa}
P.~Tribedy, R.~Venugopalan, {QCD saturation at the LHC: comparisons of models
  to p+p and A+A data and predictions for p+Pb collisions}, Phys.Lett. B710
  (2012) 125--133.
\newblock \href {http://arxiv.org/abs/1112.2445} {\path{arXiv:1112.2445}},
  \href {http://dx.doi.org/10.1016/j.physletb.2012.02.047}
  {\path{doi:10.1016/j.physletb.2012.02.047}}.

\bibitem{Kopeliovich:2005ym}
B.~Z. Kopeliovich, J.~Nemchik, I.~K. Potashnikova, M.~B. Johnson, I.~Schmidt,
  {Breakdown of QCD factorization at large Feynman x}, Phys. Rev. C72 (2005)
  054606.
\newblock \href {http://arxiv.org/abs/hep-ph/0501260}
  {\path{arXiv:hep-ph/0501260}}, \href
  {http://dx.doi.org/10.1103/PhysRevC.72.054606}
  {\path{doi:10.1103/PhysRevC.72.054606}}.

\bibitem{prep}
H.~F. J.L.~Albacete, A.~Dumitru, Y.~Nara, in preparation.

\bibitem{QuirogaArias:2010wh}
P.~Quiroga-Arias, J.~G. Milhano, U.~A. Wiedemann, {Testing nuclear parton
  distributions with pA collisions at the TeV scale}, Phys.Rev. C82 (2010)
  034903.
\newblock \href {http://arxiv.org/abs/1002.2537} {\path{arXiv:1002.2537}},
  \href {http://dx.doi.org/10.1103/PhysRevC.82.034903}
  {\path{doi:10.1103/PhysRevC.82.034903}}.

\bibitem{Dumitru:2005gt}
A.~Dumitru, A.~Hayashigaki, J.~Jalilian-Marian, {The color glass condensate and
  hadron production in the forward region}, Nucl. Phys. A765 (2006) 464--482.
\newblock \href {http://arxiv.org/abs/hep-ph/0506308}
  {\path{arXiv:hep-ph/0506308}}, \href
  {http://dx.doi.org/10.1016/j.nuclphysa.2005.11.014}
  {\path{doi:10.1016/j.nuclphysa.2005.11.014}}.

\bibitem{JalilianMarian:2011dt}
J.~Jalilian-Marian, A.~H. Rezaeian, {Hadron production in pA collisions at the
  LHC from the Color Glass Condensate}, Phys.Rev. D85 (2012) 014017.
\newblock \href {http://arxiv.org/abs/1110.2810} {\path{arXiv:1110.2810}},
  \href {http://dx.doi.org/10.1103/PhysRevD.85.014017}
  {\path{doi:10.1103/PhysRevD.85.014017}}.

\bibitem{Stasto:2011ru}
A.~Stasto, B.-W. Xiao, F.~Yuan, {Back-to-Back Correlations of Di-hadrons in dAu
  Collisions at RHIC}\href {http://arxiv.org/abs/1109.1817}
  {\path{arXiv:1109.1817}}.

\bibitem{Strikman:2010bg}
M.~Strikman, W.~Vogelsang, {Multiple parton interactions and forward double
  pion production in pp and dA scattering}, Phys. Rev. D83 (2011) 034029.
\newblock \href {http://arxiv.org/abs/1009.6123} {\path{arXiv:1009.6123}},
  \href {http://dx.doi.org/10.1103/PhysRevD.83.034029}
  {\path{doi:10.1103/PhysRevD.83.034029}}.

\bibitem{Kang:2011bp}
Z.-B. Kang, I.~Vitev, H.~Xing, {Dihadron momentum imbalance and correlations in
  d+Au collisions}, Phys.Rev. D85 (2012) 054024.
\newblock \href {http://arxiv.org/abs/1112.6021} {\path{arXiv:1112.6021}},
  \href {http://dx.doi.org/10.1103/PhysRevD.85.054024}
  {\path{doi:10.1103/PhysRevD.85.054024}}.

\bibitem{Lappi:2012xe}
T.~Lappi, H.~Mantysaari, {Forward dihadron correlations in the Gaussian
  approximation of JIMWLK}\href {http://arxiv.org/abs/1207.6920}
  {\path{arXiv:1207.6920}}.

\bibitem{JalilianMarian:2012bd}
J.~Jalilian-Marian, A.~H. Rezaeian, {Prompt photon production and photon-hadron
  correlations at RHIC and the LHC from the Color Glass Condensate}, Phys.Rev.
  D86 (2012) 034016.
\newblock \href {http://arxiv.org/abs/1204.1319} {\path{arXiv:1204.1319}},
  \href {http://dx.doi.org/10.1103/PhysRevD.86.034016}
  {\path{doi:10.1103/PhysRevD.86.034016}}.

\bibitem{Stasto:2012ru}
A.~Stasto, B.-W. Xiao, D.~Zaslavsky, {Drell-Yan Lepton-Pair-Jet Correlation in
  pA collisions}, Phys.Rev. D86 (2012) 014009.
\newblock \href {http://arxiv.org/abs/1204.4861} {\path{arXiv:1204.4861}},
  \href {http://dx.doi.org/10.1103/PhysRevD.86.014009}
  {\path{doi:10.1103/PhysRevD.86.014009}}.

\bibitem{Braidot:2010zh}
E.~Braidot, f.~t.~S. collaboration, {Suppression of Forward Pion Correlations
  in d+Au Interactions at STAR}\href {http://arxiv.org/abs/1005.2378}
  {\path{arXiv:1005.2378}}.

\end{thebibliography}







\end{document}